# Droplet beam generated with the round-tip axicon: Exact solutions for different axicon shapes

V. Yu. Mylnikov[*] and G. S. Sokolovskii

*Ioffe Institute, St. Petersburg, 194021, Russia*
*Corresponding author: vm@mail.ioffe.ru*

(Dated: February 24, 2022)

We theoretically investigate droplet quasi-Bessel beams for different shapes of the round-tip axicon. Exact solutions for the Fresnel diffraction integrals describing the axial distribution of the electric field amplitude behind the axicon are demonstrated. The analysis of the exact solutions shows that the period of "light droplets" is not a constant value, but depends on the axial distance and on the deviation of the axicon surface from the conical shape far from the rounded region. The predicted effect can be applied for the reconstruction of the exact shape of the axicon surface without 3D scanning.

## I. Introduction

In theory, an infinite plane wave will transform into an ideal Bessel beam after passing through the axicon. A beam of this type travels through space without diffraction and contains an infinite amount of energy [1]. In practice, a Bessel beam is usually formed from a Gaussian beam [2]. Light radiation behind the axicon, in this case, has a finite energy and propagation length, which depends on the aperture of the forming beam. The Bessel beam can be effectively used for optical imaging [3], particle trapping and manipulation [4], terahertz photonics [5], material processing [6], and high-contrast light sheet microscopy [7]. Following demonstrations of Bessel beam generation by semiconductor lasers and LEDs [8,9], the possibility of their practical application has grown significantly.

Recently, quasi-Bessel beams with a central core beam that propagates intermittently in the form of light droplets have been intensively studied [7,10,11]. Such droplet beams have practical applications, particularly in high-resolution microscopy [11], which requires a high-contrast image of thick opaque objects without distortion. The resolution of the droplet beam is comparable to that of a standard optical system based on a Gaussian beam. Furthermore, in contrast to a Gaussian beam, the light droplets retain their spatial profile and provide full illumination inside opaque hollow objects. Droplet beams can also be used in optogenetics to study the dynamics of living cells [12] and neuronal activity [13].

However, in experimental setups, expensive spatial light modulators are commonly used to form droplet quasi-Bessel beams [7,10,11], which significantly complicate setup adjustment and operation. Therefore, using an axicon with a round tip (Fig. 1 (a)) to generate droplet quasi-Bessel beams [14–16] is an appealing alternative that allows significant simplification and cost savings in the experimental setup.

The emergence of a droplet structure in a quasi-Bessel beam formed by a round-tip axicon is associated with differences in wave vector projections onto the symmetry axis for components of light passing through different regions of the axicon. The central part of the radiation, passing through the rounded region, which acts as a focusing lens maintains the projection of the wave vector onto the symmetry axis. The rest of the light radiation is refracted at the conical region of the axicon and thus receives a smaller projection of the wave vector, which is determined by the angle of refraction [15–17]. Interference between these parts of radiation causes oscillations in the axial intensity distribution. Their half-period is equal to $\Delta z = \lambda/2[1-\cos(\gamma)] \approx \lambda/\gamma^2$ (for $\gamma \ll 1$) [17], where $\lambda$ is a light wavelength, and $\gamma$ is the propagation angle of light after refraction on the axicon's conical surface (Fig. 1(a)). This explanation seems to be universal and independent of the detailed shape of the axicon rounded tip.

In this paper, we theoretically investigate droplet quasi-Bessel beams for different shapes of the round-tip axicon. Exact solutions for the Fresnel diffraction integrals describing the axial distribution of the electric field amplitude behind the axicon are demonstrated. From the analysis of the obtained solutions, we conclude that the interference structure of the droplet quasi-Bessel beam is significantly influenced not only by the shape of the round tip but also by the deviation of the axicon surface from the conical shape far from the rounded region. The main observable consequence is that the oscillation period of light droplets is dependent on the distance and shape of the axicon surface. The predicted effect can be applied for the reconstruction of the exact shape of the axicon surface without 3D scanning.

## II. Round-tip axicon models

The experimental setup is depicted schematically in Fig. 1(a). Assume the axicon is illuminated by a Gaussian beam with the following electric field amplitude:

$$E_0(r) = \exp\left(-\frac{r^2}{2w_0^2}\right), \qquad (1)$$

where $r = (x^2+y^2)^{1/2}$ is a transverse plane radial coordinate, and $w_0$ is the beam waist. The axicon's propagation is accounted for by multiplying the electric field amplitude (1) by the phase mask $U(r)$ of the form:

$$U(r) = \exp[-ik(n-1)f(r)], \qquad (2)$$

where $k = 2\pi/\lambda$ is a vacuum wavenumber, $\lambda$ is a wavelength, $n$ is the refractive index, and $f(r)$ is a function that defines the axicon surface shape.

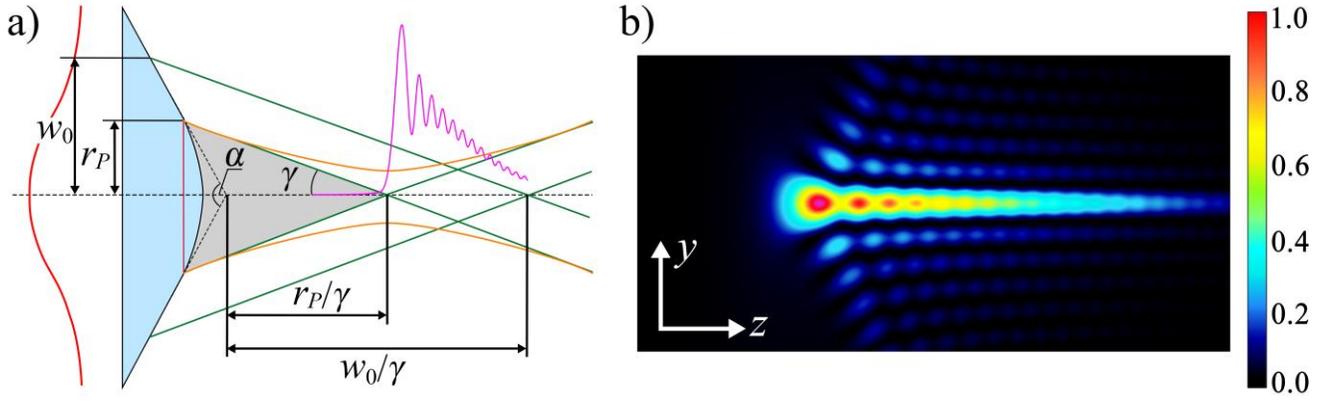

Fig. 1. (a) A Gaussian beam with a waist radius of $w_0$ is fed into a round-tip axicon with a semi-apex angle $\alpha$ and radius of the rounded region $r_P$. The light radiation transmitted through the conical part of the axicon propagates at an angle $\gamma$ with respect to the $z$ axis (green lines), forming a Bessel beam near the axis. The distances $r_P/\gamma$ and $w_0/\gamma$ define the beginning and end of the Bessel beam's geometric propagation region. The radiation is focused at a point with the coordinate $z=r_P/\gamma$ by the round tip of the axicon acting as a thin lens (orange curve). The interference of "conical" and "near-spherical" field components causes oscillations in the axial intensity distribution (purple curve). Grey paint depicts the geometrical shadow area near the round tip. (b) A quasi-Bessel beam's axial intensity evolution behind the axicon. The colorbar is shown on the right of the figure, and the color from red to black represents the intensity from high to low.

A conical surface of the form denotes the function $f(r)$ far from the round tip:

$$f(r) \approx \upsilon r, \quad (3)$$

Where $\upsilon=\text{ctg}(\alpha/2)$ is the coefficient preceding the linear function $r$, which depends on the axicon's apex angle. Light after the axicon propagates at an angle of $\gamma=\upsilon(n-1)$ with respect to the $z$ direction in this case. Radiation close to the axis produces a Bessel beam with a maximum propagation length $z_{max}=w_0\text{ctg}(\gamma)\approx w_0/\gamma$ (for $\gamma\ll1$) (Fig. 1(a)).

However, near the rounded region the function $f(r)$ should be approximated by a parabola:

$$f(r) \approx \upsilon\left(r_D + \frac{r^2}{2r_P}\right), \quad (4)$$

where $r_P$ is a radius of the rounded region or the curvature radius of the parabolic function multiplied by $\upsilon$ at $r=0$, and $r_D$ is the distance from the round to the ideal tip of the axicon divided by $\upsilon$. As a result of equation (4), the round tip of the axicon acts on light as a thin lens with a focal length equal to $F=r_P/\upsilon(n-1)$.

Fig. 1 depicts the droplet structure of the central core of a quasi-Bessel beam formed by a round-tip axicon. The interference of radiation passing through a "thin lens" and the conical part of the axicon causes it to appear [15–17]. The droplet oscillations' half-period is equal to ($\gamma\ll1$):

$$\Delta z = \lambda / 2(1-\cos(\gamma)) \approx \lambda / \gamma^2. \quad (5)$$

According to equation (5), the half-period is determined solely by the refraction angle $\gamma$ and is unaffected by the shape of the round tip. We also show that the half-period of oscillations is not a constant value, but is determined by the shape of the axicon surface and depends on the distance to the round tip.

To determine the surface properties that can affect the droplet structure of a quasi-Bessel beam, we consider three models of the axicon shape. All these models are described by the conical surface (3) far from the round tip and by the paraboloid (4) near it. The first of these models is given by a hyperbolic surface of the form [17]:

$$f_1(r) = \upsilon\sqrt{r_P^2 + r^2}. \quad (6)$$

The distance parameter $r_D$ is no longer an independent variable in this case and is equal to $r_D=r_P$. According to equation (6), the hyperbolic function near the tip ($r\ll r_P$) behaves like a parabola (4), but far from the tip ($r\gg r_P$) it is described by a linear function (3) and a small correction term:

$$f_1(r) = \begin{cases} \upsilon\left(r_P + \dfrac{r^2}{2r_P}\right), & r \ll r_P \\ \upsilon r + \upsilon r_P^2/2r, & r \gg r_P \end{cases}. \quad (7)$$

The second model is given by the parabola and cone [16]:

$$f_2(r) = \begin{cases} \upsilon\left(\dfrac{r_P}{2} + \dfrac{r^2}{2r_P}\right), & r \leq r_P \\ \upsilon r, & r > r_P \end{cases}. \quad (8)$$

The first obvious difference between these two models is a difference in the value of the parameter $r_D$. It is equal to $r_D=r_P$ for the hyperbolic model, but it is less than that for the parabolic-cone model, $r_D=r_P/2$. This difference is clearly shown in Fig. 2, where the rounded regions for both models are the same, but the values of the height parameter $r_D$ differ.

However, as will be demonstrated below, the most significant difference between the considered models is the behavior of the surfaces in the region far from the round tip. Assume that $r\gg r_P$ and the following asymptotic relations can be obtained:

$$f_1(r) \approx \upsilon r + \upsilon r_P^2/2r, \quad (9)$$

$$f_2(r) = \upsilon r, \qquad (10)$$

where $\delta f_1(r)=[f_1(r)-\upsilon r]\approx \upsilon r_P^2/2r$ is the deviation of the hyperbolic surface (6) from the case of an ideal cone (3) far from the round tip. As a result of equation (10), the deviation for the parabolic-cone model (8) is strictly zero. These considerations demonstrate that the hyperbolic and parabolic-cone models are not equivalent and should result in different intensity distributions behind the axicon, as shown below.

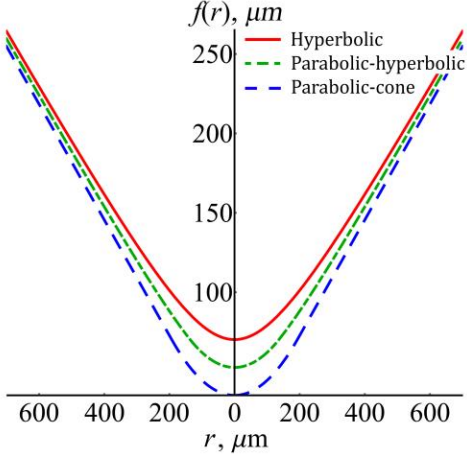

Fig. 2. An axicon surface functions $f(r)$ for the hyperbolic model (red solid curve), determined by equation (6), the parabolic-cone model (blue dashed curve), determined by equation (8), and the parabolic-hyperbolic model for $\varepsilon=0.7$, determined by equation (11). Numerical simulations were run with $r_P=190$ μm and $\alpha=140°$ ($\upsilon=0.36$).

To track the changes in the transition from one surface shape to another, one must build a function $f(r)$, which yields equation (6) in the first limit and equation (8) in the second. In order to accomplish this, we propose a parabolic-hyperbolic model, which has the following form:

$$f_3(r) = \begin{cases} \upsilon\left(\dfrac{r_P}{2}(1+\varepsilon^2) + \dfrac{r^2}{2r_P}\right), & r \le r_P\sqrt{1-\varepsilon^2} \\ \upsilon\sqrt{\varepsilon^2 r_p^2 + r^2}, & r > r_P\sqrt{1-\varepsilon^2} \end{cases}, \qquad (11)$$

where $\varepsilon$ is a control parameter in parabolic-hyperbolic model, which accepts values from 0 to 1. When $\varepsilon=0$, the hyperbolic part transforms into a cone, and equation (11) becomes (8). When $\varepsilon=1$, the surface (11) coincides with the hyperbolic one (6). Thus, the main difference between the considered parabolic-hyperbolic model and the well-known hyperbolic [15,17], parabolic-cone [14], and sphere-cone [16] models, as well as the hyperbolic-cone model [18], is the presence of two independent parameters instead of one.

Furthermore, the asymptotic form of the considered parabolic-hyperbolic model (11) far from the round tip ($r>>r_P$) is as follows:

$$f_3(r) \approx \upsilon r + \upsilon\varepsilon^2 r_P^2/2r. \qquad (12)$$

According to equation (12) switching from the parabolic-cone model (8) to the hyperbolic one (6) provides control over the deviation of the axicon surface from the case of the ideal cone $\delta f(r)=[f(r)-\upsilon r]\approx \upsilon\varepsilon^2 r_P^2/2$. As will be shown below, the variation of the parameter $\varepsilon$ affects the period of intensity oscillations associated with the droplet beam structure formed by an axicon with a round tip.

## III. The Hyperbolic model

Consider how the change from one model of the axicon surface to another affects the oscillation structure of the axial intensity distribution. The Fresnel diffraction integral determines the axial evolution of a droplet quasi-Bessel beam [19]. In the following, we will consider only the axial ($r=0$) electric field distribution behind the axicon, which will be equal to in the paraxial approximation:

$$E(z) = \frac{k}{iz}\int_0^\infty dr\, r\exp\left[-\left(\frac{1}{w_0^2}-i\frac{k}{z}\right)\frac{r^2}{2}-ik(n-1)f(r)\right]. \qquad (13)$$

It should also be noted that in Eq. (13) we ignore phase contribution $\exp(ikz)$. For the hyperbolic model (6) the axial electric field amplitude (13) takes the following form [17]:

$$E(\xi) = \frac{g}{i\xi}\int_0^\infty d\rho\, \rho\exp\left[-\left(1-i\frac{g}{\xi}\right)\frac{\rho^2}{2}-ig\sqrt{\xi_P^2+\rho^2}\right]. \qquad (14)$$

where new $w_0$-scaled variables are introduced as follows:

$$\xi = \frac{\gamma}{w_0}Z, \quad \xi_P = \frac{r_P}{w_0}, \quad g = kw_0\gamma, \qquad (15)$$

where $\xi$ is a normalized axial coordinate, $\xi_P$ is a normalized curvature radius of the rounded tip, and $g$ is a rescaled refraction angle. It turns out that integral (14) can be evaluated analytically. It is worth noting that no analytical expression has been obtained in the papers on round-tip axicons that we are aware of [15,17,18,20–24]. After changing the integration variable to $u=(\xi_P^2+\rho^2)^{1/2}$, the exact evaluation can be performed.

$$E(\xi) = \frac{g}{i\xi}e^{\left(1-i\frac{g}{\xi}\right)\frac{\xi_P^2}{2}}\int_{\xi_P}^\infty du\, u\exp\left[-\left(1-i\frac{g}{\xi}\right)\frac{u^2}{2}-igu\right]. \qquad (16)$$

As a result, Eq. (18) is a well-known integral [25], which is expressed in terms of the complementary error function erfc(x) [26] as follows:

$$\int_c^\infty du\, u\exp\left[-\frac{au^2}{2}-ibu\right] = \frac{1}{a}\exp\left[-ibc-\frac{ac^2}{2}\right] - \sqrt{\frac{\pi}{2}}\frac{ib}{a^{3/2}}\exp\left[-\frac{b^2}{2a}\right]\text{erfc}\left(\frac{ac+ib}{\sqrt{2a}}\right). \qquad (17)$$

So, the exact solution for the axial electric field amplitude can be written as:

$$E(\xi) = \frac{e^{-ig\xi_P}}{1+i\xi/g} + \exp\left(-\frac{ig}{2}\frac{\left[\xi^2+(1+i\xi/g)^2\xi_P^2\right]}{\xi(1+i\xi/g)}\right) \times \\ \times \frac{\sqrt{-i\pi g\xi/2}}{(1+i\xi/g)^{3/2}}\text{erfc}\left(\sqrt{\frac{-ig}{2\xi}}\frac{(\xi_P-\xi+i\xi_P\xi/g)}{\sqrt{1+i\xi/g}}\right). \qquad (18)$$

We are interested in the regime $g \gg 1$, for which Eq. (18) can be greatly simplified. For example, for an axicon with $\alpha = 140°$ and $n=1.5$ ($\gamma = 0.18$) and radiation at the wavelength of $\lambda = 1$ μm with beam waist $w_0=175$ μm the normalized refraction angle is $g=200$. As a result, in the arguments of the exponent and complementary error function, the first nonvanishing correction term proportional to the small parameter $1/g$ must be considered. Thus, the exponent argument will be as follows:

$$-\frac{ig}{2} \frac{\left[\xi^2 + (1+i\xi/g)^2 \xi_P^2\right]}{\xi(1+i\xi/g)} \approx$$
$$\approx -\frac{\xi^2 - \xi_P^2}{2} - \frac{ig(\xi^2 + \xi_P^2)}{2\xi}. \quad (19)$$

The argument to the complementary error function can be written in the form:

$$\sqrt{\frac{-ig}{2\xi}} \frac{(\xi_P - \xi + i\xi_P \xi/g)}{\sqrt{1+i\xi/g}} \approx$$
$$\approx \sqrt{\frac{-ig}{2\xi}} \left[(\xi_P - \xi) + \frac{i\xi}{g}(\xi_P + \xi)\right]. \quad (20)$$

Using (19) and (20), the electric field amplitude can be expressed as follows:

$$E(\xi) = \exp(-ig\xi_P) + \exp\left(-\frac{\xi^2}{2} + \frac{\xi_P^2}{2} - \frac{ig}{2}\left(\xi + \frac{\xi_P^2}{\xi}\right)\right) \times$$
$$\times \sqrt{\frac{-i\pi g\xi}{2}} \operatorname{erfc}\left(\sqrt{\frac{-ig}{2\xi}}\left[(\xi_P - \xi) + \frac{i\xi(\xi_P + \xi)}{g}\right]\right). \quad (21)$$

The complementary error function must now be simplified. To that end, it makes sense the expanding the error function $\operatorname{erfc}(\tau + \delta\tau)$ as a Taylor series in small perturbations of the parameter $\delta\tau$ up to the first order:

$$\operatorname{erfc}(\tau + \delta\tau) \approx \operatorname{erfc}(\tau) + \delta\tau \operatorname{erfc}^{(1)}(\tau), \quad (22)$$

where $\operatorname{erfc}^{(1)}(\tau)$ is the first derivative of a complementary error function, $\tau=(-ig/2\xi)^{1/2}(\xi_P-\xi)$ and $\delta\tau=(i\xi/2g)^{1/2}(\xi_P+\xi)$, which will be proportional to $\tau \propto g^{1/2}$, and $\delta\tau \propto 1/g^{1/2}$. Then we have the following relationship:

$$\delta\tau \operatorname{erfc}^{(1)}(\tau) \propto 1/g^{1/2}, \quad (23)$$

because the first derivative of $\operatorname{erfc}^{(1)}(\tau) \propto 1$. The second derivative of the error function, on the other hand, increases linearly with $\tau$ for $\tau \gg 1$, and we have:

$$\delta\tau^2 \operatorname{erfc}^{(2)}(\tau) \propto \delta\tau^2 \tau \propto 1/g^{1/2}. \quad (24)$$

In other words, the second term in the Taylor series is of the same order of smallness as the first. As a result, considering all contributions proportional to $1/g^{1/2}$, we must explicitly sum the infinity Taylor series. In our case, this can be done for $\tau \gg 1$. The $k^{th}$-order derivative of the error function can be expressed in this limit as follows:

$$\operatorname{erfc}^{(k)}(\tau) \approx \frac{2}{\sqrt{\pi}} \exp(-\tau^2)(-2\tau)^{k-1}. \quad (25)$$

After using $\tau \gg 1$ and (25), the Taylor series become:

$$\sum_{k=1}^{\infty} \frac{\delta\tau^k}{k!} \operatorname{erfc}^{(k)}(\tau) \approx -\frac{\exp(-\tau^2)}{\sqrt{\pi}\tau} \sum_{k=1}^{\infty} \frac{(-2\tau\delta\tau)^k}{k!} =$$
$$= \frac{\exp(-\tau^2)}{\sqrt{\pi}\tau} \left[1 - \exp(-2\tau\delta\tau)\right]. \quad (26)$$

Then (26) and (21) result in:

$$E(\xi) = \frac{\xi e^{-\frac{\xi^2}{2} + \frac{\xi_P^2}{2}} - \xi_P}{\xi - \xi_P} e^{-ig\xi_P} + \sqrt{\frac{-i\pi g\xi}{2}} \times$$
$$\times \exp\left(-\frac{\xi^2}{2} + \frac{\xi_P^2}{2} - \frac{ig}{2}\left(\xi + \frac{\xi_P^2}{\xi}\right)\right) \operatorname{erfc}\left(\sqrt{\frac{-ig}{2\xi}}(\xi_P - \xi)\right). \quad (27)$$

It may appear that an Eq. (27) has a divergent term at the point $\xi = \xi_P$. However, it is clear that:

$$\lim_{\xi \to \xi_P} \frac{\xi_P - \xi e^{-\frac{\xi^2 - \xi_P^2}{2}}}{\xi - \xi_P} = \xi_P^2 - 1. \quad (28)$$

Furthermore, it is convenient to replace the complex argument's complementary error function with the well-known Fresnel integrals [26]:

$$\operatorname{erfc}\left(\sqrt{-i}x\right) = 1 - \sqrt{\frac{2}{i}}\left[C\left(\sqrt{\frac{2}{\pi}}x\right) + iS\left(\sqrt{\frac{2}{\pi}}x\right)\right], \quad (29)$$

which are defined as:

$$C(x) = \int_0^x \cos\left(\frac{\pi t^2}{2}\right) dt, \quad S(x) = \int_0^x \sin\left(\frac{\pi t^2}{2}\right) dt. \quad (30)$$

As a result, we get the following final expression:

$$E(\xi) = \frac{e^{-ig\xi_P}}{\xi - \xi_P}\left[\xi \exp\left(-\frac{\xi^2}{2} + \frac{\xi_P^2}{2}\right) - \xi_P\right] +$$
$$+ \sqrt{\pi g\xi} \exp\left(-\frac{\xi^2}{2} + \frac{\xi_P^2}{2} - \frac{ig}{2}\left(\xi + \frac{\xi_P^2}{\xi}\right)\right) \times \quad (31)$$
$$\times \left[\frac{1}{2} + S\left(\sqrt{\frac{g}{\pi\xi}}(\xi - \xi_P)\right) - i\left(\frac{1}{2} + C\left(\sqrt{\frac{g}{\pi\xi}}(\xi - \xi_P)\right)\right)\right].$$

As shown in Fig. 3(a-b), the simplified expression (31) almost exactly coincides with the rigorous field amplitude determined by the formula (18). This confirms correctness of asymptotic expansion of Eq. (18) over a small parameter $1/g$.

Figure 4(a-b) depicts the calculated axial intensity distributions $I(\xi)=|E(\xi)|^2$ of the droplet quasi-Bessel beam. For numerical simulation, we used simplified expression (31). As previously stated. The round tip of the axicon produces axial intensity distribution oscillations [17], which are caused by the interference of light passing through the "thin lens" and the conical region of the axicon.

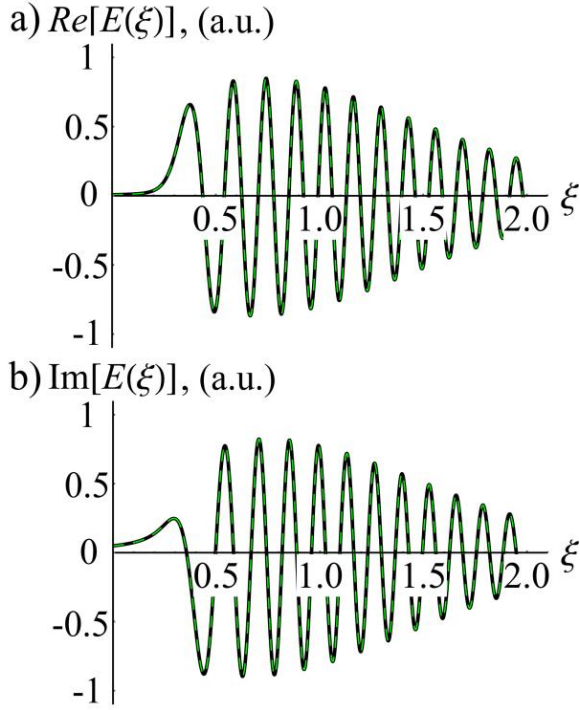

Fig. 3. The exact formula (18) (black thick curve) and its simplified version (31) (green dashed curve) are used to calculate the axial distributions of the real (a) and imaginary (b) parts of the electric field amplitude. Numerical simulations use $g=100$ and $\xi_P=0.3$.

However, the received expression (31) cannot be represented by the sum of two interfering components. Here, we modify it so that the resulting expression only contains the Bessel beam and a small correction from a "thin lens." For this, the asymptotes of the Fresnel integrals for $x\gg1$ should be used:

$$C(x) \approx -\frac{1}{2} + \theta(x) + \frac{1}{\pi x}\sin\left(\frac{\pi}{2}x^2\right),$$
$$S(x) \approx -\frac{1}{2} + \theta(x) - \frac{1}{\pi x}\cos\left(\frac{\pi}{2}x^2\right), \quad (32)$$

where $\theta(x)$ is a Heaviside step function. When we substitute (32) into the expression (31), we get:

$$E(\xi) = \sqrt{2\pi g \xi}\exp\left(-\frac{\xi^2}{2} + \frac{\xi_P^2}{2} - \frac{ig}{2}\left(\xi + \frac{\xi_P^2}{\xi}\right) - i\frac{\pi}{4}\right) \times$$
$$\times \theta(\xi - \xi_P) - \frac{\xi_P}{\xi - \xi_P}\exp(-ig\xi_P). \quad (33)$$

As a result, the first term in expression (33) describes a Bessel beam that is constrained by the Heaviside step function. Its occurrence is due to the presence of a geometrical shadow area originating from the round tip that prevents formation of the Bessel beam, as shown in grey in Fig. 1(a). This region's presence results in a significant suppression of the axial intensity ($r=0$) for $\xi<\xi_P$, as seen in Fig. 4(a). The second term refers to the radiation component that passes through the round tip or "thin lens." As a result of the interference of a Bessel beam and a convergent nearly-spherical wave, oscillations of the axial intensity distribution emerge. It is also worth noting that the spherical wave term in Eq. (33) diverges at the point $\xi=\xi_P$.

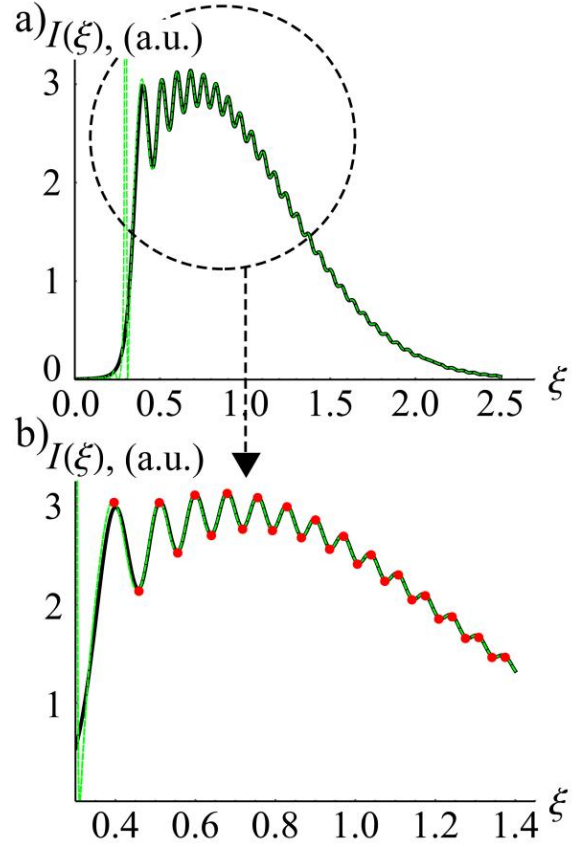

Fig. 4. Axial intensity distribution $I(\xi)$ of the droplet quasi-Bessel beam behind the axicon calculated with simplified formula (31) shown by black solid line and with its approximation (36) shown by green dashed line (a). A closer look at the axial intensity oscillations (b). The red dots represent the extremum points of the intensity distribution calculated using Eqs. (38) and (39). Numerical simulations are run with $g=200$ and $\xi_P=0.3$.

An expression very similar to (33) was previously derived phenomenologically [16]. The primary distinction between formula [16] and (33) above is the Bessel beam phase factor. Let us examine it further by returning to dimensional coordinates and restoring the phase factor $\exp(ikz)$:

$$E_{BB}(z) \propto \exp\left[ik\left(1-\frac{\gamma^2}{2}\right)z - ik\frac{r_P^2}{2z}\right]. \quad (34)$$

The first phase term proportional to $k(1-\gamma^2/2)z\approx k\cos(\gamma)z$ results from refraction of radiation on the "conical" region of the axicon surface and subsequent propagation at an angle $\gamma$ with respect to the $z$ direction. The presence of the second phase term $\exp[-ikr_P^2/2z]$ distinguishes (34) from the corresponding expression in [16].

To obtain a simple physical explanation for this contribution, we will consider the geometrical optics approach. As shown in Fig. 1(a), light refracted on the conical region of the axicon at a point with radius $r$ propagates at an angle $\gamma$ with respect to the axis of symmetry to the point $z=r/\gamma$. In this case, the phase factor $\exp[-ik(n-1)\delta f(r)]$, which arises from the axicon surface's deviation from the conical surface, is also transferred from a point with radius $r$ in the axicon plane to the axis of symmetry $z=r/\gamma$. As a result, the Bessel beam gains an additional phase factor that is explicitly dependent on the distance $z$ and the round tip parameter $r_P$:

$$E_{BB}(z) \propto \exp\left[-ik\frac{r_P^2}{2z}\right] = \exp\left[-\frac{ig}{2}\frac{\xi_P^2}{\xi}\right]. \quad (35)$$

Furthermore, we will show that the arising phase factor (35) causes the oscillation period to be dependent on the distance $z$ from the axicon round tip. Eq. (33) provides an explicit expression for the axial intensity distribution in this case:

$$I(\xi) = \left[\sqrt{2\pi g \xi}\exp\left(-\frac{\xi^2}{2}+\frac{\xi_P^2}{2}\right)\theta(\xi-\xi_P) - \right.$$
$$\left. -\frac{\xi_P}{\xi-\xi_P}\cos\left(\frac{g}{2}\left(\xi+\frac{\xi_P^2}{\xi}\right)+\frac{\pi}{4}-g\,\xi_P\right)\right]^2, \quad (36)$$

where the first term in square brackets makes the main contribution to the intensity and the second term containing the cosine function is responsible for the oscillations. In this case, position of the extremums can be found as points where the argument of the cosine is a multiple of $\pi$. Sorting the contributions in descending order of powers of $\xi$, we obtain the following quadratic equation for the extremum points:

$$\xi^2 - \xi_0(m)\xi + \xi_P^2 = 0, \quad (37)$$

where we've used the notation $\xi_0(m)=2\pi(m-1/4)/g+2\xi_P$. The solution to equation (37) is as follows:

$$\xi(m) = \frac{1}{2}\left(\xi_0(m) + \sqrt{\xi_0(m)^2 - 4\xi_P^2}\right). \quad (38)$$

The extremum points of the axial intensity distribution determined by expression (38) are indicated by red dots in Fig. 4(b). Furthermore, because the radical expression in (38) becomes negative for $m<1$, the first extremum appears at $m=1$. It also follows from (38) that $\xi(m)\to\xi_0(m)$ for $m\gg1$. The distance between the two closest extremum points that can be interpreted as a half-period of oscillations is given by:

$$\Delta\xi(m) = \xi(m+1) - \xi(m), \quad (39)$$

and shown in Fig. 5(a). In this figure, index $m$ denotes the positions of extremum points of the axial intensity distribution. As a result, increasing index $m$ in Fig. 5(a) corresponds to increasing coordinate $\xi$ in Fig. 4(b).

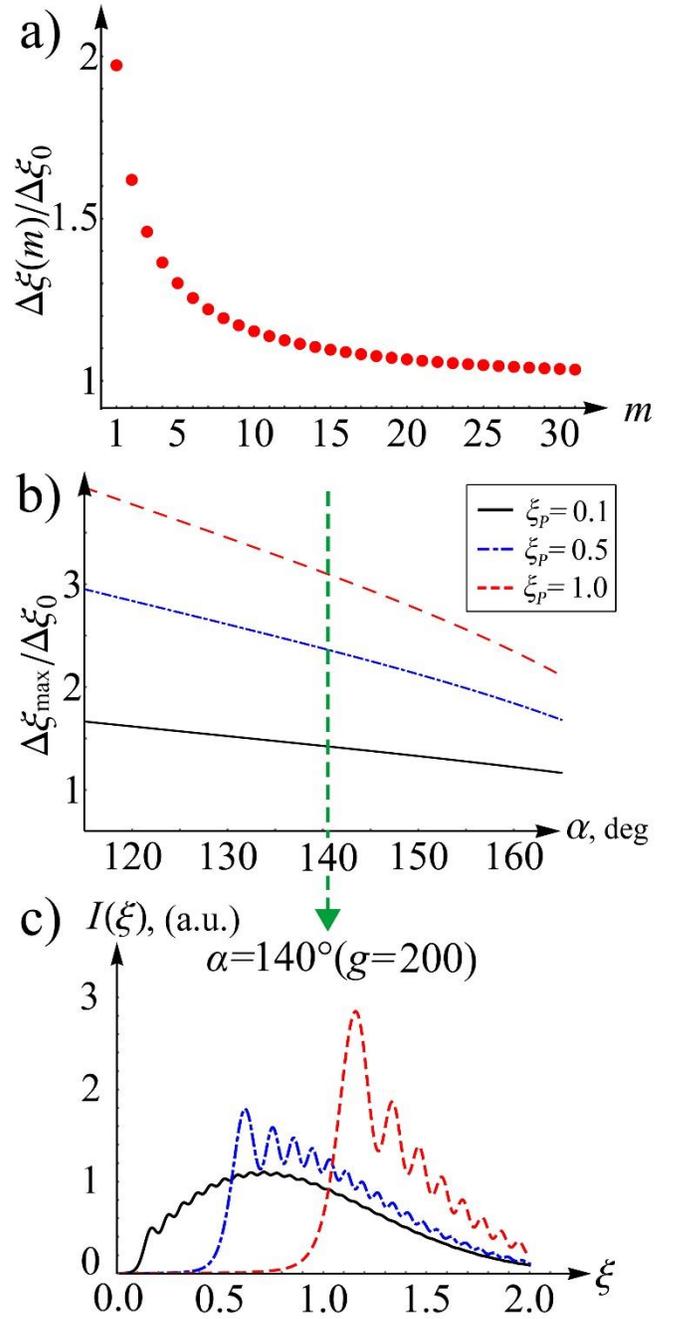

Fig. 5. (a) Normalized half-period of the intensity oscillations $\Delta\xi(m)/\Delta\xi_0$ vs the index m of the extremum points (red dots) for $g=200$ and $\xi_P=0.3$. The half-period $\Delta\xi(m)$ is given by Eq. (39) and normalized by the minimal half-period $\Delta\xi_0=2\pi/g$. (b) Normalized maximal half-period $\Delta\xi_{max}/\Delta\xi_0$ vs apex angle $\alpha$ for various $\xi_P$ values. (c) The axial intensity distributions $I(\xi)$ for $\lambda=1$ μm, $n=1.5$, $w_0=175$ μm and $\xi_P=0.1$ (black solid), $\xi_P=0.5$ (blue dash-dotted), $\xi_P=1.0$ (red dotted).

Figs. 5(a-c) shows the half-period of the oscillations to be dependent on the extremum index $m$ (and, hence, on the coordinate $\xi$) as well as on the shape of the round-tip axicon surface. The last follows from the fact that additional phase (35) is proportional to the curvature radius $r_P^2$ ($\xi_P^2$ in dimensionless coordinates). For $m\gg1$ ($\xi\gg1$), the half-period $\Delta\xi(m)$ tends to a constant value $\Delta\xi_0=2\pi/g$

that follows from the interference considerations [17]. This is due to the properties of the additional phase factor (35), which is proportional to $1/\xi$ and decreases with increasing distance $\xi$ as clearly illustrated in Figs. 5(b) and 5(c), where the half-period of oscillations of the axial intensity distribution increases as $\xi_P$ increases.

Thus, for $m>>1$ additional phase factor (35) contributes to the total phase and the half-period $\Delta\xi(m)$ less than the phase resulting from the light propagation at an angle $\gamma$ with respect to the symmetry axis (Fig. 1(a)). However, for small $m$, the half-period $\Delta\xi(m)$ can significantly deviate from its minimal value $2\pi/g$. For instance, the maximum half-period $\Delta\xi(m=1)$ is nearly twice its minimal value $\Delta\xi_0=2\pi/g$ for $\xi_P=0.3$ and $g=200$. Fig. 5(b) shows the dependence of the maximal oscillation half-period $\Delta\xi_{max}=\Delta\xi(m=1)$ on the apex angle $\alpha$ for fixed values of $\xi_P=0.1$ (black solid), $\xi_P=0.5$ (blue dash-dotted), $\xi_P=1.0$ (red dotted). One can see that lower angle $\alpha$ and higher $\xi_P$ lead to higher values of the maximal half-period.

As a result, we've shown that the half-period of oscillations for the hyperbolic surface model is not constant and is affected by the axial coordinate $\xi$ and the shape of the axicon surface. Following that, we will demonstrate what will change in the case of the parabolic-hyperbolic model and specifically for the parabolic-cone model.

## IV. The parabolic-hyperbolic and parabolic-cone models

Let us now look at the parabolic-hyperbolic model, which is defined by equation (11). The electric field amplitude determined by the Fresnel diffraction integral (13) takes the following form:

$$E(\xi) = E_P(\xi) + E_H(\xi), \quad (40)$$

$$E_P(\xi) = \frac{g}{i\xi}\int_0^{\xi_P\sqrt{1-\varepsilon^2}} d\rho\,\rho\exp\left[-\left(1-i\frac{g}{\xi}\right)\frac{\rho^2}{2} - ig\left(\xi_P\frac{1+\varepsilon^2}{2} + \frac{\rho^2}{2\xi_P}\right)\right], \quad (41)$$

$$E_H(\xi) = \frac{g}{i\xi}\int_{\xi_P\sqrt{1-\varepsilon^2}}^{\infty} d\rho\,\rho\exp\left[-\left(1-i\frac{g}{\xi}\right)\frac{\rho^2}{2} - ig\sqrt{\varepsilon^2\xi_P^2+\rho^2}\right], \quad (42)$$

where $E_P(\xi)$ and $E_H(\xi)$ are the electric field contributions from the parabolic and hyperbolic regions of the axicon surface (11), respectively. By changing the integration variable to $u=(\varepsilon^2\xi_P^2+\rho^2)^{1/2}$, integral (41) can be evaluated.

$$E_H(\xi) = \frac{g}{i\xi}e^{\frac{1}{2}\left(1-i\frac{g}{\xi}\right)\varepsilon^2\xi_P^2}\int_{\xi_P}^{\infty} du\,u\exp\left[-\left(1-i\frac{g}{\xi}\right)\frac{u^2}{2} - igu\right], \quad (43)$$

The only difference between (43) and (16) is the pre-integral exponent with factor containing $\varepsilon\xi_P$. Using this fact and assuming $g>>1$ we get in analogy to (31):

$$E_H(\xi) = \frac{\xi e^{-\frac{\xi^2-\xi_P^2}{2}} - \xi_P}{\xi-\xi_P}\exp\left[-ig\xi_P - \frac{\xi_P^2(1-\varepsilon^2)}{2}\left(1-\frac{ig}{\xi}\right)\right] +$$

$$+\sqrt{\pi g\xi}\exp\left(-\frac{\xi^2-\varepsilon^2\xi_P^2}{2} - \frac{ig}{2}\left(\xi+\frac{\varepsilon^2\xi_P^2}{\xi}\right)\right)\times \quad (44)$$

$$\times\left[\frac{1}{2} + S\left(\sqrt{\frac{g}{\pi\xi}}(\xi-\xi_P)\right) - i\left(\frac{1}{2} + C\left(\sqrt{\frac{g}{\pi\xi}}(\xi-\xi_P)\right)\right)\right].$$

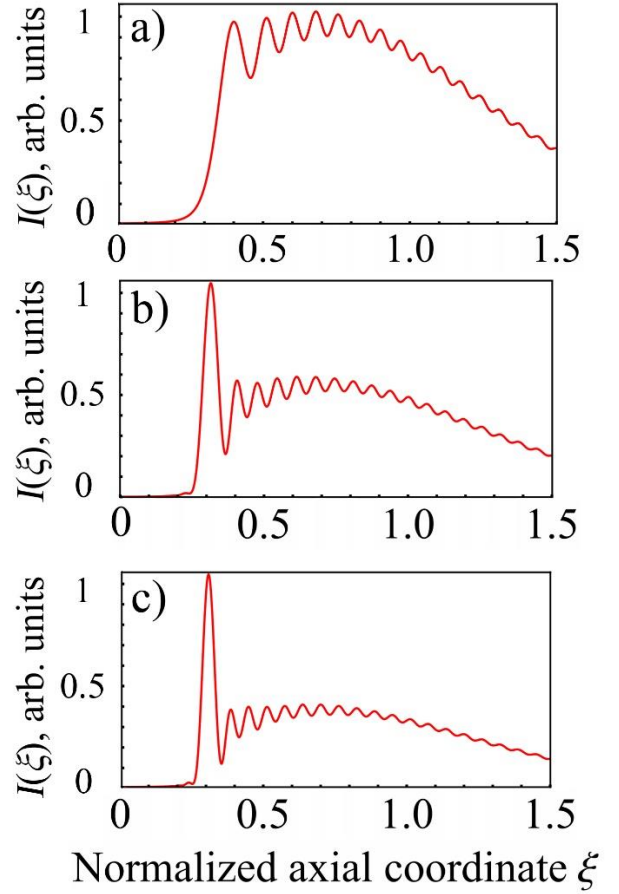

Fig. 6. The axial intensity distributions $I(\xi)$ calculated with Eq. (46) for $g=200$, $\xi_P=0.3$ and the following parameters: (a) $\varepsilon=1$ (hyperbolic model), (b) $\varepsilon=0.5$ (parabolic-hyperbolic model), and (d) $\varepsilon=0$ (parabolic-cone model).

The integral (41) corresponding to the parabolic region of the axicon can also be evaluated explicitly, and for $g>>1$ we obtain:

$$E_P(\xi) = -\frac{\xi_P}{\xi-\xi_P-i\xi_P^2/g}\left[\exp\left(-ig\sqrt{1-\varepsilon^2}\,\xi_P\right) - \right.$$

$$\left. -\exp\left(-\frac{\xi_P^2(1-\varepsilon^2)}{2}\left(1-\frac{ig}{\xi}\right) - ig\xi_P\right)\right]. \quad (45)$$

The total amplitude of the electric field is given by the sum of contributions from the parabolic (45) and hyperbolic (44) parts of the surface:

$$E(\xi) = \frac{e^{-ig\xi_P}}{\xi - \xi_P} \left( \xi e^{-\frac{\xi^2}{2} + \frac{\varepsilon^2 \xi_P^2}{2} + ig\frac{\xi_P^2(1-\varepsilon^2)}{2\xi}} - \xi_P e^{ig\frac{\xi_P^2(1-\varepsilon^2)}{2\xi_P}} \right) +$$

$$+ \sqrt{\pi g \xi} \exp\left( \frac{\varepsilon^2 \xi_P^2 - \xi^2}{2} - \frac{ig}{2}\left( \xi + \frac{\varepsilon^2 \xi_P^2}{\xi} \right) \right) \times \quad (46)$$

$$\times \left[ \frac{1}{2} + S\left( \sqrt{\frac{g}{\pi \xi}} (\xi - \xi_P) \right) - i \left( \frac{1}{2} + C\left( \sqrt{\frac{g}{\pi \xi}} (\xi - \xi_P) \right) \right) \right].$$

Formula (46) unifies all considered models of the axicon surface with parameter $\varepsilon$ responsible for transition from the hyperbolic model for $\varepsilon=0$ (see Fig. 6(a)) to the parabolic-cone model for $\varepsilon=1$ (Fig. 6(c)) through the parabolic-hyperbolic model for all other values $0<\varepsilon<1$ (Fig. 6(b)). One can see that the transition from the hyperbolic to the parabolic-cone model results in a significant increase in the first peak of the intensity oscillations, as shown in the Fig. 6.

Let us simplify the electric field amplitude by rewriting Eq. (46) as the sum of the Bessel and near-spherical waves, similar to (33):

$$E(\xi) = \sqrt{2\pi g \xi} \exp\left( -\frac{\xi^2}{2} + \frac{\varepsilon^2 \xi_P^2}{2} - \frac{ig}{2}\left( \xi + \frac{\varepsilon^2 \xi_P^2}{\xi} \right) - i\frac{\pi}{4} \right) \times \quad (47)$$

$$\times \theta(\xi - \xi_P) - \frac{\xi_P}{\xi - \xi_P} \exp\left( -ig\xi_P \frac{(1+\varepsilon^2)}{2} \right).$$

From here, the axial intensity will be as follows:

$$I(\xi) = \left[ \sqrt{2\pi g \xi} \exp\left( -\frac{\xi^2}{2} + \frac{\varepsilon^2 \xi_P^2}{2} \right) \theta(\xi - \xi_P) - \right.$$

$$\left. -\frac{\xi_P}{\xi - \xi_P} \cos\left( \frac{g}{2}\left( \xi + \frac{\varepsilon^2 \xi_P^2}{\xi} \right) + \frac{\pi}{4} - g\xi_P \frac{(1+\varepsilon^2)}{2} \right) \right]^2. \quad (48)$$

The following conclusions are drawn from the analysis of the expressions for the axial electric field amplitude (47) and intensity (48):

(i) The edge of the geometrical shadow is determined by the point $\xi_P = r_P/w_0$. This position remains constant throughout the transition from one model of the axicon surface to another, which is consistent with the intuitive geometrical optics consideration.

(ii) The amplitude factor $\exp(\varepsilon^2\xi_P^2/2)$ of the Bessel beam in expression (47) is responsible for the significant increase of the first peak of oscillations at the point $\xi \approx \xi_P$ in Fig. 6. The influence of this factor is obviously the highest for the hyperbolic model ($\varepsilon=1$) and decreases to unity for the parabolic-cone model ($\varepsilon=0$). This results in smaller contribution of the Bessel beam to the total axial intensity compared to that of a near-spherical wave with transition from the hyperbolic to the parabolic-cone model of the axicon.

iii) The additional phase factor of the Bessel beam in normalized/dimensional coordinates will take the following form:

$$E_{BB}(\xi) \propto \exp\left[ -\frac{ig}{2} \frac{\varepsilon^2 \xi_P^2}{\xi} \right] = \exp\left[ -ik \frac{\varepsilon^2 r_P^2}{2z} \right]. \quad (49)$$

As a result, it is determined by $\varepsilon r_P$, rather then $r_P$, as was the case for the hyperbolic model of the axicon surface (35). With (49), equation determining the axial intensity distribution's extremum points (38) will change as follows:

$$\xi(m) = \frac{1}{2}\left( \xi_0(m) + \sqrt{\xi_0(m)^2 - 4\varepsilon^2 \xi_P^2} \right), \quad (50)$$

where $\xi_0(m) = 2\pi(m-1/4)/g + \xi_P(1+\varepsilon^2)$.

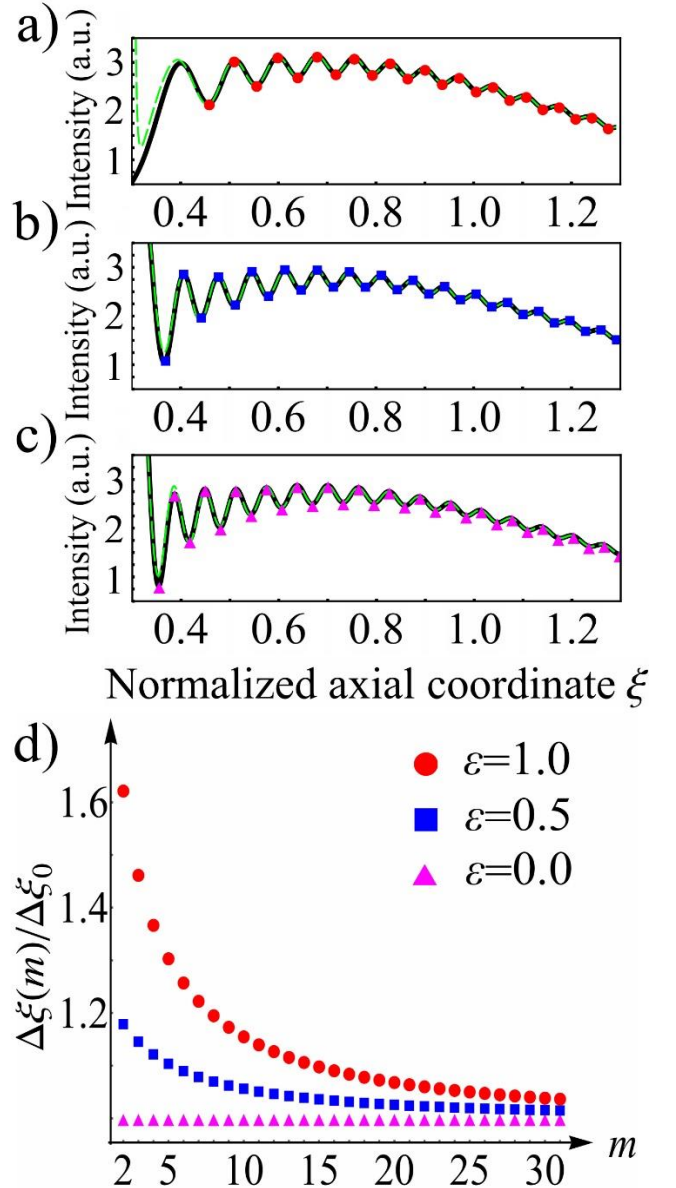

Fig. 7. Axial intensity distributions $I(\xi)$ of a droplet quasi-Bessel beam behind an axicon with normalized refraction angle $g=200$ and curvature radius $\xi_P=0.3$ calculated using (46) (black curve) and its approximation (48) (green dashed curve) for $\varepsilon=1$ (a), $\varepsilon=0.5$ (b) and $\varepsilon=0$ (c). Indicators in (a-c) denote the local maxima and minima points of the intensity distribution calculated with Eqs. (50) and (39) and serve to derive the corresponding normalized half-period of the intensity oscillations $\Delta\xi(m)$ vs the maxima/minima index $m$ (d).

According to the analysis of the phase factor (50) and Figs. 7(a, d), the hyperbolic model of the axicon surface has the strongest dependence of the period of "light droplets" on distance. This is due to the highest value of parameter $\varepsilon=1$ in this model. Dependence of the "light droplets" period on the coordinate $z$ is far less pronounced for the parabolic-hyperbolic model, as shown in Figs. 7(b,d), due to a decrease in the parameter $\varepsilon$. Finally, this dependence vanishes for the parabolic-cone model (Figs. 7(c,d)) because the phase in expression (52) vanishes itself. As a result, we have shown that the axial intensity distribution of the droplet quasi-Bessel beam is influenced not only by the shape of the axicon round tip, which is defined by the parameter $r_P$ in the models under consideration. As the transition from the hyperbolic to the parabolic-cone model weakens the dependence of the period of "light droplets" on the coordinate $z$ (or index $m$ in Fig. 7(d)), it is clear that the deviation of the axicon surface from the conical shape far from the rounded tip is also very important. This is taken into account by the parameter $\varepsilon$, which sheds the light on the differences between very similar hyperbolic [17] and parabolic-cone [16] models of the axicon and on the effect of the axicon surface shape far from the rounded tip.

The results presented in this paper can be used to reconstruct the shape of the axicon surface without any need of 3D scanning. First of all, the parameter $r_P$ determining the curvature of a rounded tip can be found in contactless experiment by measuring the amplitude of the oscillations or the position of the geometrical shadow. If contactless measurement isn't required, this parameter can be easily determined from a profilometer cross-section of the axicon surface. However, the second independent parameter $\varepsilon$ cannot be determined from the same cross section as the profile measurement through the axicon center is not guaranteed. However, parameter $\varepsilon$ can be found in contactless experiment through the measurement of the axial dependence of the period of the "light-droplets" of the quasi-Bessel beam. This can be done using formulae (50) and (39) to fit the experimental data with the known rounded tip curvature $r_P$.

## V. CONCLUSION

In this paper, we study the properties of droplet quasi-Bessel beams formed by the round-tip axicon. We consider the generalized shape of the axicon surface, which is determined by a parabolic-hyperbolic model with well-known hyperbolic [17] and parabolic-cone [16] models being its special cases. We show that for all considered surface shapes, exact analytical solutions for the axial distribution of the electric field amplitude behind the axicon can be obtained. The analysis of the exact solutions shows that the period of "light droplets" is not a constant value, but depends on the axial distance and on the shape of the axicon surface. We demonstrate that the transition from the hyperbolic to the parabolic-hyperbolic model weakens axial dependence of the period of "light droplets", and for the parabolic-cone model this dependence vanishes completely. Also, we demonstrate that transition from one model to another result in a significant increase in the relative amplitude of the first peak of the light intensity oscillations. As a result, not only does the axicon round tip have a significant effect on the droplet structure of a quasi-Bessel beam, but so does the deviation of the axicon surface from the conical shape far from the rounded region. The practical significance of this study stems from the ability to reconstruct the shape of the axicon surface contactless without the use of 3D scanning. Furthermore, the obtained results may be very useful for design of cost-effective optics for generation of controlled axial intensity distributions [27] including bottle beams [28].